\documentclass[manuscript]{aastex}
\usepackage{multirow}

\begin{document}

\title{The contribution of X-ray binaries to the evolution
of late-type galaxies: Evolutionary population synthesis
simulations}
\author{Zhao-Yu Zuo$^{1,2,3}$ and Xiang-Dong Li$^{1,3}$}
\affil{$^1$Department of Astronomy, Nanjing
University, Nanjing 210093, China;\\
$^2$Department of Optical Information Sciences and Technology,
School of Science, Xi'an Jiaotong
University, Xi'an 710049, China\\
$^3$Key laboratory of Modern Astronomy and Astrophysics (Nanjing
University), Ministry of Education, Nanjing 210093,
China\\zuozyu@mail.xjtu.edu.cn; lixd@nju.edu.cn}

\begin{abstract}

X-ray studies of normal late-type galaxies have shown that
non-nuclear X-ray emission is typically dominated
by X-ray binaries, and provides a useful measure of star formation activity.
We have modeled the X-ray evolution of
late-type galaxies over the $\sim$ 14 Gyr of cosmic history, with
an evolutionary population synthesis code developed by Hurley et
al. Our calculations reveal a decrease of the X-ray
luminosity-to-mass ratio $L_{\rm X}/M$ with time, in agreement
with observations (Fig.~7$a$). We show that this decrease is a
natural consequence of stellar and binary evolution and mass
accumulating process in galaxies. The  X-ray-to-optical luminosity
ratio $L_{\rm X}/L_{\rm B}$ is found to be fairly constant (around
$\sim 10^{30}$ erg\,s$^{-1}$$L_{\rm B,\odot}^{-1}$,
Fig.~7$b$), and insensitive to the star formation history in the
galaxies. The nearly constant value of $L_{\rm X}/L_{\rm B}$
is in conflict with the observed
increase in $L_{\rm X}/L_{\rm B}$ from $z=0$ to
1.4. The discrepancy may be caused by intense obscured star
formation activity that leads to nonlinear relationship between X-ray
and B-band emission.

\end{abstract}


\section{Introduction}
The X-ray emission of a normal late-type galaxy (i.e., one without
an active galactic nuclei) is often dominated by the integrated
emission of the galactic X-ray binaries (XRBs)
\citep[e.g.,][]{fabbiano03,colbert04,fabbiano06}. Galactic XRBs
can be classified into two distinct populations \citep{bhattacharya91}: the
short-lived ($\lesssim 10^6$ yr), high-mass X-ray binaries (HMXBs)
and the long-lived ($> 10^8$ yr), low-mass X-ray binaries (LMXBs).
The X-ray emission from HMXBs is usually regarded to  trace
current star formation because of their short lifetimes,
while the X-ray emission from LMXBs is more closely related to the
integrated stellar mass \citep{ptak01,ran03,grimm03}.

With the observations of galaxies at redshift $z>0.1$, either
from deep surveys by {\it Chandra\/} and {\it XMM-Newton\/}
\citep[see][for a review]{brandt05} or from stacking analysis of
distant galaxy fields \citep[$z\simeq 0.1-4$;
e.g.,][]{brandt01,hornschemeier02,nandra02,georgakakis03,reddy04,laird05,laird06,lehmer05,lehmer08},
it has become possible to investigate the X-ray properties of normal
galaxies at cosmologically significant redshifts
\citep{hornschemeier00,hornschemeier03,alexander02,georgantopoulos05,georgakakis07,lehmer06,lehmer07,kim06,Tzanavaris06,rosa07}.
Previous studies showed that the average X-ray luminosities
$L_{\rm X}$  of normal late-type galaxies increase with redshift
out to $z\simeq 1.4-3$, and evolve as $(1+z)^{1.5-3}$ over the
redshift range $z\simeq 0-1.4$ for X-ray-detected normal galaxies
\citep{norman04,ptak07,tzanavaris08}. \citet{hornschemeier02} performed
a statistical X-ray study of spiral galaxies in the {\it Hubble} Deep Field-North
and its flanking fields using the {\it Chandra} Deep Field-North 1 Ms data set,
and observed a factor of
$\sim 2-3$ increase in the X-ray-to-optical luminosity ratio
$L_{\rm X}/L_{\rm B}$ from $z=0$ to 1.4 for the $L_{\rm
B}$-selected galaxies. To improve the constraints on the X-ray
evolution of late-type galaxies, \citet{lehmer08} studied
for the first time how the X-ray properties evolve as a function
of optical luminosity, stellar mass, and star formation rate (SFR)
of the  galaxies in the {\it Chandra\/} Deep Fields
North and South \citep{alexander03,giacconi02}. It was found that
there is a significant increase (by a factor of about $5-10$) in
the X-ray-to-optical luminosity ratio ($L_{\rm X}/L_{\rm B}$) and
the X-ray-to-stellar mass ratio ($L_{\rm X}/M$) for the galaxy
populations selected by $L_{\rm B}$ and $M$, respectively, over
the redshift range of $z=0-1.4$. When analyzing the galaxy samples
selected with SFR, these authors found that the X-ray
luminosity-to-SFR ratio ($L_{\rm X}$/SFR) is constant over the
entire redshift range for galaxies with SFR $= 1-100 M_{\odot}\rm
yr^{-1}$, and that the star formation activity (as traced by X-ray
luminosity) per unit stellar mass in a given redshift bin
increases with decreasing stellar mass over the redshift range $z
= 0.2-1$, consistent with previous studies on how star formation
activity depends on stellar mass
\citep{cowie96,juneau05,bundy06,noeske07a,noeske07b,zheng07}.
Finally, they extended their X-ray analyses to Lyman break
galaxies at $z\sim$ 3 and estimated that the value of $L_{\rm X}/L_{\rm B}$ at
$z\sim 3$ is similar to that at $z=1.4$.

X-ray emission of normal galaxies and its evolution have also been the
subject of theoretical studies. Using a semi-empirical approach to
link XRB lifetimes with a cosmological evolution of SFR,
\citet{ghosh01} discussed the imprints left by this cosmic SFR on
the evolution of X-ray luminosities $L_{\rm X}$ of normal
galaxies. They showed that the evolving SFRs can strongly affect
the integrated galactic X-ray emission, with the possibility of
significant evolution of the X-ray luminosities even within relatively
low redshifts $z< 1$ \citep[see also][]{white98}.
\citet{eracleous04} simulated the evolution of X-ray luminosities
of XRBs after a burst of star formation with duration of 20 Myr with a
population synthesis method, and found that the $2-10$ keV
luminosity reaches a maximum after approximately 20 Myr, and the
X-ray luminous phase can be sustained for a period of hundreds of
Myr. The results were shown to be insensitive to the initial mass
function (IMF) and the average mass ratio between accreting and
donor stars. However, a comprehensive study on the evolution of
X-ray populations in galaxies and their relation with other
properties is still lacking.

In the present work, we use an evolutionary population synthesis
(EPS) code to calculate the X-ray luminosity of XRBs and its
evolution in a normal late-type galaxy over the cosmic history.
Meanwhile, we calculate the optical luminosity and the galactic
mass contributed by stellar populations.  The objective of this
study is to investigate the X-ray evolution of late-type
galaxy populations, its dependence on the physical properties of
galaxies (e.g., optical luminosity, stellar mass and mass-to-light
ratio) and on the star formation history (SFH), from a theoretical
point of view. We will also examine how the key parameters, such
as IMF, common envelope (CE) efficiency, the binary fraction and
metallicity, may affect the X-ray emission of the galaxies.
In Section 2 we introduce the method of calculation and model
parameters. In Section 3 the calculated results are presented
and compared with observations. We summarize in Section 4.
Throughout the paper, we assume a flat $\Lambda$ cold dark matter
($\Lambda$CDM) cosmology with $\Lambda_{\rm m,0}=0.3$,
$\Omega_{\Lambda,0}=0.7$, and $H_0=70$ kms$^{-1}$Mpc$^{-1}$
\citep{spergel03}, which imply a look-back time of 7.7 Gyr at
$z=1$.

\section{Models}

\subsection{Assumptions and input parameters}

We used the EPS code developed by \citet{Hurley00,Hurley02} and
updated by \citet[][see Appendix A in the paper]{liu07} and
\citet{zuo08} to calculate the X-ray luminosity $L_{\rm X}$ of
XRBs, the optical luminosity $L_{\rm B}$ and the stellar mass $M$
of the host galaxy, as well as their evolution. The values of the
adopted parameters are the same as the default ones in
\citet{Hurley02} if not mentioned otherwise.

Previous works \citep{Shapley01,persic07,lehmer08} have already
shown that
the galactic X-ray emission is closely related to the SF activity. So we
constructed three cases, i.e., constant SF, star-burst
SF and cosmic SF cases, to examine their effect. We also examined
several key parameters, such as the IMF, CE efficiency parameter, the binary
fraction and metallicity (listed in Table~1 and discussed below)
to explore their influence on the X-ray evolution of the galaxies.

\noindent\emph{1. Constant star formation case}

In this case, we adopt a constant SFR $=0.25 M_{\odot}$yr$^{-1}$
for stars more massive than $5 M_{\odot}$ derived by \citet{grimm03} in
our Galaxy, and take the star formation duration (SFD) to be
14 Gyr. For each model, we evolve $10^6$ primordial binary
systems, with the same grid of initial parameters (i.e., primary
and secondary mass, orbital separation) as Hurley et al. (2002).
We also evolve $10^6$ primordial single stars, with initial mass
logarithmically spaced between $0.1 M_{\odot}$ and $80 M_{\odot}$.
In our basic model (i.e., model M1, listed in Table~1) we assume
the binary fraction $f$ to be 0.5 and evolve each binary and
single star on the grid. In the following we describe the
assumptions and input parameters in our basic model.

In order to be in parallel with \citet{lehmer08}, we take the IMF
of \citet[][hereafter KROUPA01]{Kroupa01} for the mass ($M_1$)
distribution of the primary stars. For the secondary stars (of mass $M_2$)
and binary orbit, we assume a uniform distribution
between 0 and 1 for the mass ratio $q\equiv M_2/M_1$, and a uniform
distribution for the logarithm of the orbital separation $\ln a$
(Hurley et al. 2002). We fix the metallicity to be
solar over the lifetime of the simulated galaxy.

We assume that any system entering Roche-lobe overflow
becomes circularized and
synchronized by tidal interaction between the binary components
(Belczynski et al. 2008).
An important parameter in the binary evolution is the CE efficiency
parameter
$\alpha_{\rm CE}$ \citep{paczynski76,iben93}, which describes the
efficiency of converting orbital energy into the kinetic energy
ejecting the envelope \citep[see \S 2.1.1 in][for detail]{zuo10}.
It can often reduce the orbital separation of the surviving
binaries by a factor of $\sim 100$, resulting in different
outcomes of binary systems. In our basic model, we adopt
$\alpha_{\rm CE}=0.3$, which can best model the luminosity function of the galaxy (Zuo et al. 2008).

We also construct several other models (listed in Table~1) by
varying the key input parameters described as follows.

(1) As stated above, variations of
the CE parameter can considerably change the relative numbers of
XRBs. However reliable values of $\alpha_{\rm CE}$ are
difficult to estimate due to lack of understanding of the processes
involved, although in the literature it is in the range from $\sim$ 0.1 to $\sim$
3.0 (e.g., Taam \& Bodenheimer 1989; Tutukov \& Yungelon 1993;
Podsiadlowski, Rappaport \& Han 2003). Here we also adopt
$\alpha_{\rm CE}=1.0$ (Model M2) to examine its effect. 

(2) The IMF determines the percentage of high-mass
stars, consequently the number of XRBs produced, and the
X-ray luminosity of the galaxy. So we also make use of the IMF of
\citet[][hereafter KTG, model M5]{Kroupa}, which is much steeper in
the high-mass end than in KROUPA01. For the secondary masses ($M_2$),
we assume the mass ratio $q$ follows
a power-law distribution $P(q)\propto q^{\alpha}$, and adopt
both the
conventional choice of flat mass spectrum, i.e., $\alpha=0$
\citep[][our basic model, M1]{Mazeh92,goldberg94,Shatsky02}
and $\alpha=1$ (Model M3), since recent data are more
consistent with ``twins" being a general feature of the
close-binary population \citep{dalton95,kobulnicky07}.

(3) Surveys of
M dwarfs within 20 pc from the Sun indicate that the binary
fraction $f$ may be a function of stellar spectral types
\citep{fischer92}, for example, $f>0.5$ for G stars and $f>0.6$
for massive O/B stars in the Cygnus OB2 association
\citep{lada06,kobulnicky07}. So we also adopt $f=0.8$ (Model M4)
for comparison.

(4) Observations of the hosts of XRBs revealed that XRBs, especially
ultra-luminous X-ray sources (ULXs) may prefer to occur in
galaxies with low metallicities \citep{mapelli09}.
So we vary the metallicity to examine its effect
on the X-ray luminosity evolution by taking $Z=1.5 Z_{\odot}$
(Model M6), $0.5 Z_{\odot}$ (Model M7), $0.1 Z_{\odot}$ (Model M8)
and $0.02 Z_{\odot}$ (Model M9), in order to compare with our
basic model (M1 with $Z_{\odot}$). The other
parameters in these models are the same as the ones in our basic
model.

\noindent\emph{2.  Star-burst star-formation case}

In galaxies like our own Galaxy, continuous star formation processes
may last for several Gyr, but it is not the case for star-burst
galaxies. For example, the Antennae galaxies may have experienced
SF for the last several
hundred Myr with an enhanced SFR of $7.1M_{\odot}$yr$^{-1}$
\citep{barnes88,mihos93}. To reveal the effect of SFH we take
the following combinations of SFD and metallicities: 100
Myr/$Z_{\odot}$ (Model M10), 20 Myr/$Z_{\odot}$ (Model M11) and
100 Myr/0.02$Z_{\odot}$ (Model M12). We assume that the SF is quenched
after the SFD time and set other parameters to be the same as
in our basic model.

\noindent\emph{3. Cosmic star-formation case}

With improved observations of star formation processes, cosmic SFH can be
constrained quite tightly within $\sim 30\%-50\%$ up to $z\sim$1
and within a factor of $\sim 3$ up to $z\sim 6$ \citep{hopkins04},
which makes it possible to investigate the cosmic X-ray evolution
of galaxies. Here we adopt the derived expression of the SFH in
\citet{hopkins06},
\begin{eqnarray}
\dot{\rho}_{\rm SF}(z)&\propto&\left\{
\begin{array} { ll}
  (1+z)^{3.44}&\ \ z\leq0.97 \\
  (1+z)^{-0.26}&\ \ 0.97\leq z\leq4.48 \\
  (1+z)^{-7.8}&\ \ 4.48\leq z, \\
\end{array}
\right.
\end{eqnarray}
and scale the SFR at redshift $z=0$ to be the same as that of our
Galaxy. Moreover, it is known that the cosmic metallicity also
evolves strongly with redshift, and galaxies at higher redshift
tend to have lower metallicities
\citep{pettini99,prochaska03,rao03,kobulnicky04,kewley07,
kulkarni05,kulkarni07,savaglio05,savaglio09,savaglio06,wolf07,peroux07}.
So we adopt an empirical equation $Z/Z_{\odot}\propto
10^{-\gamma z}$ \citep{langer06} for metallicity evolution
with $\gamma=0.15$ \citep{kewley07}. We also vary the IMF
to a steeper one (KTG93) and a shallower one \citep[][BG03 for
short]{baldry03} to examine its effect in this case. The other
parameters are the same as in our basic model.

\subsection{X-ray luminosity and source type}
We adopt the same procedure to calculate the $0.5-8$ keV X-ray
luminosities of different XRB populations as in \citet{zuo10}.
Mass transfer in XRBs occurs via either Roche-lobe overflow or
capture of the wind material from the donor star. We use the
classical \citet{Bondi44}'s formula to calculate the wind
accretion rate of the compact stars. In the case of Roche-lobe overflow,
mass is transferred to the accreting star by way of an accretion disk. It
is known that accretion disks in LMXBs are subject to the thermal instability
if the accretion rate is sufficiently low \citep{paradijs96}. We
discriminate transient and persistent LMXBs according to the
criteria of \citet{paradijs96}  for main sequence (MS) and red
giant donors, and of \citet{ivanova06} for white dwarf (WD)
donors, respectively. The simulated X-ray luminosity is described as
follows,
\begin{eqnarray}
L_{\rm X, 0.5-8 keV}&=&\left\{
\begin{array} { ll}
  \eta_{\rm bol}\eta_{\rm out}L_{\rm Edd}&\ \rm transients\ in\ outbursts, \\
  \eta_{\rm bol}\min(L_{\rm bol},\eta_{\rm Edd}L_{\rm Edd})&\ \rm persistent\
  systems,
\end{array}
\right.
\end{eqnarray}
where the bolometric accretion luminosity $L_{\rm bol}\simeq
0.1\dot{M}_{\rm acc}c^2$ (where $\dot{M}_{\rm acc}$ is the
accretion rate and $c$ is the velocity of light), the critical
Eddington luminosity $L_{\rm Edd} \simeq 4\pi Gm_{1}m_{\rm
p}c/\sigma_{\rm T}=1.3 \times 10^{38}m_{1}$\,ergs$^{-1}$ (where
$\sigma_{\rm T}$ is the Thomson cross section, $m_{\rm p}$ the
proton mass, $G$ the gravitational constant, and $m_{1}$ the
accretor mass in the units of solar mass), and $\eta_{\rm Edd}$ is
the factor to allow super-Eddington luminosities, taken to be
5 \citep{ohsuga02,beg02}. To transform the bolometric luminosity into the
$0.5-8$ keV X-ray luminosity, a bolometric correction factor
$\eta_{\rm bol}$ is introduced \citep{bel04}. Generally, its
value is $\sim 0.1-0.5$ for different types of XRB, here we adopt
$\eta_{\rm bol}\simeq 0.1$. For transient sources the X-ray
luminosity during outbursts should be larger than the long-term
one by a factor $\eta_{\rm out}$. We take $\eta_{\rm out}=0.1$ and
1 for the short and long-period systems, and the
critical periods are adopted to be 1 day for neutron star (NS)
transients and 10 hours for black hole (BH) transients,
respectively \citep{chen97,Garcia03,bel08}.

\subsection{Optical luminosity $L_{\rm B}$ and stellar mass of the galaxy}

The optical luminosity $L_{\rm B}$ of a galaxy is mostly
from normal stars (both binary and single stars).
Assume that the
stellar radiation can be reasonably approximated as a blackbody,
the B-band luminosity of a star is calculated with $L_{\rm
B}=\frac{\int_{\rm B}I_{\lambda} d \lambda}{\int I_{\lambda} d
\lambda} \times L$, where $L$ is the total thermal luminosity of
the star, and the radiative intensity
$I_{\lambda}=\frac{2hc^2}{\lambda^5}\frac{1}{e^{hc/\lambda kT_{\rm
eff}}-1}$, where $h$ is the Planck constant, $k$ the Boltzmann
constant, and $T_{\rm eff}$ the effective temperature.

We also examine the contribution of optical luminosity
from accretion disks in XRBs,
resulting from the reprocessing of X-ray photons.
We calculate the optical
luminosity $L_{\rm B}$ from the accretion disk in BH XRBs following
\citet{madhusudhan08}, adopting the same
temperature profile (i.e., their Eq.~[4]) to describe the effective
temperature in the disk.  For NS XRBs we find similar results as BH XRBs.
Our calculation reveals that optical radiation from accretion disks in XRBs is
negligible compared to the overall stellar optical luminosity.

The stellar mass here  is the sum of the masses of currently living stars,
and does not include the contribution from
compact stars (WDs, NSs, and BHs),
in order to be in parallel with \citet{lehmer08},
where they used the rest-frame $B-V$ color and $K$-band luminosity to
estimate the masses of the galaxies.

\section{Results}

\subsection{Constant star-formation case}

Figure~1 shows the calculated values of $L_{\rm X}$, $L_{\rm B}$,
$L_{\rm X}/M$, $L_{\rm X}/L_{\rm B}$, $L_{\rm X}/(M/L_{\rm B})$
and $M/L_{\rm B}$ against time in the constant SF
case. The five panels from left to right correspond to the basic
model (M1), models with $\alpha_{\rm CE}=1.0$ (M2), $\alpha=1$
(M3), $f=0.8$ (M4), and KTG93 IMF (M5), respectively.

For X-ray luminosities $L_{\rm X}$ (Fig.~1$a$) we plot the
contributions from HMXBs and LMXBs with dotted and dashed lines, respectively.
The X-ray luminosity of HMXBs rises rapidly shortly after the first SF,
and remains nearly constant afterward, because of continuous,
constant SF. LMXBs take much longer time to form,
and their number is correlated with the total star mass, resulting in
a long-term increasing trend in X-ray luminosity with time. The position of the
crossing point of the two lines depends most strongly on the CE parameter
$\alpha_{\rm CE}$ (larger $\alpha_{\rm CE}$ results in more
LMXBs).

The optical luminosity $L_{\rm B}$ (Fig.~1$b$) is
contributed by the primary stars (dotted line) and secondary stars (dashed
line) in binaries, and by single stars (long-dashed line).
When the fraction of binaries is larger (Model
M4), more XRBs are produced, leading to larger $L_{\rm X}$, and hence
larger $L_{\rm X}/M$, $L_{\rm X}/L_{\rm B}$ and $L_{\rm
X}/(M/L_{\rm B})$ ratios compared with those in the basic model.
The steeper end of IMF (i.e., Model M5) implies a smaller number
of massive stars (and smaller $L_{\rm B}$), resulting in fewer compact objects
and smaller $L_{\rm X}$. So the
$L_{\rm X}/M$ and $L_{\rm X}/(M/L_{\rm B})$ ratios in this case
are smaller than those in models M1-M4.

The $L_{\rm X}/M$ ratio (Fig.~1$c$) has a clear decreasing trend after
the age of tens of Myr. Observationally, this phenomena was regarded
as the evidence  that lower-mass galaxies appear to have higher specific
star formation rates than massive ones (Brinchmann et al. 2004; Bauer et al.
2005; Noeske et al. 2007a; Feulner et al. 2005; Zheng et al.
2007). Our results show that the decrease of $L_{\rm
X}/M$ is a natural consequence of XRB evolution and stellar
mass accumulating process in galaxies - the stellar mass always steadily
increases while the X-ray luminosity changes little during most of the evolution.

The $L_{\rm X}/L_{\rm B}$ ratio (Fig.~1$d$), similar to $L_{\rm
X}$, rises in the first several Myr, then remains nearly
constant afterward. The flattening values are all around $10^{30}$
erg\,s$^{-1}$$L_{\rm B,\odot}^{-1}$, however have severalfold
changes among different models. They are caused by the diversity
in the percentages of both total massive stars and massive stars in binaries,
which
determine $L_{\rm B}$ and $L_{\rm X}$, respectively.
The  ratios of the peak values of $\log (L_{\rm X}/L_{\rm B})$
are roughly $1:2:0.5:1.5:1$ from models M1 to M5.

Figure~2 shows the cumulative X-ray luminosity functions (XLFs) of
HMXBs (dotted line) and LMXBs (dashed line) in the top panels. The
parameters are the same as in Figure~1. Note that HMXBs and LMXBs
dominate at relatively high ($>10^{39}$ ergs$^{-1}$) and low
($<10^{39}$ ergs$^{-1}$) luminosity ends, respectively. We also
show the detailed components of XRBs which contribute the total
X-ray luminosity separately in the middle (HMXBs) and bottom
(LMXBs) panels of Figure~2. The dotted, dashed, dash-dotted and
dash-dot-dotted lines represent persistent BH XRBs (BHp), transient BH XRBs
(BHt), persistent NS XRBs (NSp) and transient NS XRBs
(NSt), respectively.  It is seen that for HMXBs persistent BH XRBs
contribute most to the XLF; for LMXBs, BH-XRBs (both BHp and BHt)
dominate the high luminosity end of XLF, while in the low luminosity
end transient NS-XRBs play a more important role.

Figure~3 shows the evolution of $L_{\rm X}$, $L_{\rm B}$, $L_{\rm
X}/M$, $L_{\rm X}/L_{\rm B}$, $L_{\rm X}/(M/L_{\rm B})$ and
$M/L_{\rm B}$  with metallicities taken to be $1.5
Z_{\odot}$, $Z_{\odot}$, $0.5 Z_{\odot}$, $0.1 Z_{\odot}$ and
$0.02 Z_{\odot}$, corresponding to models M6, M1, M7, M8 and M9 from left to
right, respectively. Note that both $L_{\rm X}$ and $L_{\rm B}$
have a roughly increasing trend with decreasing metallicity.
The values of $L_{\rm X}/M$ and $L_{\rm X}/L_{\rm B}$ are comparable among
different models. The values of the peak values are $\sim 1:1:0.7:1:2$ for
$\log (L_{\rm X}/M)$ and $\sim 1:1:0.5:1:1$ for $\log (L_{\rm X}/L_{\rm B})$.
The corresponding cumulative XLF are shown in
Fig.~4.


\subsection{Star-burst star-formation case}

We present the evolution of $L_{\rm X}$, $L_{\rm B}$, $L_{\rm
X}/M$, $L_{\rm X}/L_{\rm B}$, $L_{\rm X}/(M/L_{\rm B})$ and
$M/L_{\rm B}$ in the star-burst case in Fig.~5. Here the
metallicity and SFD are assumed to be $Z_{\odot}$/100 Myr (left),
$Z_{\odot}$/20 Myr (middle), and 0.02 $Z_{\odot}$/100 Myr (right).
It is seen that the X-ray luminosity $L_{\rm X}$ (Fig.~5$a$) rises
to its peak within the SF episodes (peaked at the age of
about 100 Myr for models M10 and M12, and at about 20 Myr for model
M11), then decreases with time, lasting at least several $10^8$
yr with $L_{\rm X}>10^{37}$ ergs$^{-1}$.
Note that the small serrations emerging in late evolution is
mainly caused by LMXBs (i.e., X-ray transient outbursts due to
thermal instability in the accretion disks). The optical luminosity
$L_{\rm B}$ (Fig.~5$b$) decreases sharply when the SF
process stops, since massive stars contribute significantly during
the star burst episode. The $L_{\rm X}/M$ (Fig.~5$c$) (and also
$L_{\rm X}/(M/L_{\rm B})$, Fig.~5$e$) ratio roughly follows the
trend of $L_{\rm X}$. However the slope in this case is much
steeper than in Figs.~1 and 3, because of the rapid decay of
the X-ray luminosity. The $L_{\rm X}/L_{\rm B}$ (Fig.~5$d$) ratio
in the three models are all comparable with those in Figs.~1 and
3, implying that it is intrinsically not sensitive to the
SFH of the galaxies.

\subsection{Cosmic star-formation case}

Figure~6 shows the evolution of  $L_{\rm X}$, $L_{\rm B}$, $L_{\rm
X}/M$, $L_{\rm X}/L_{\rm B}$, $L_{\rm X}/(M/L_{\rm B})$ and
$M/L_{\rm B}$ with time (left) and redshift $z$ (right), with a
cosmic SFH \citep[from][]{hopkins06} and a cosmic metallicity
evolution history \citep[from][]{langer06} taken into account. Note
that the X-ray luminosity $L_{\rm X}$ is mainly dominated by HMXBs
in this case over the whole cosmic history because of the enhanced
SFR with increasing redshift as a whole. This is comparable with the
work of \citet{lehmer08}, where they found that LMXBs on
average play a fairly small role in the X-ray emission. The $L_{\rm
X}/M$ ratio has a decreasing trend after the age of tens of Myr (or
increases with $z$), similar as in the constant/burst SF cases. This
also confirms our previous conclusion that the decrease of $L_{\rm
X}/M$ with time results from XRB evolution and
stellar mass growth in the galaxies.
The $L_{\rm X}/L_{\rm B}$ ratio rises rapidly in the first several
Myr, then stays around $10^{30}$ erg\,s$^{-1}$$L_{\rm B,\odot}^{-1}$
afterward, similar as in the constant/burst SF cases, indicating
that the $L_{\rm X}/L_{\rm B}$ ratio is not sensitive to the SFH in
the galaxies.

To compare the theoretical predictions with observations, we re-plot
Fig.~6{\it c} and 6{\it d} in Fig.~7{\it a} and 7{\it b}, with redshift
$z$ ranging from 0 to 2.0. The solid, dotted and dashed lines
represent the modeled results with IMFs of Kroupa (2001, KROUPA01),
Kroupa, Tout,
\& Gilmore (1993, KTG93) and Baldry \& Glazebrook (2003, BG03),
respectively. The observational data of $\log(L_{\rm X}/M)$ and
$\log(L_{\rm X}/L_{\rm B})$ are taken from Shapley et al. (2001, S01)
for normal late-type galaxies in the local universe (open symbols),
\citet[][L08]{lehmer08} with the stellar masses
$10^{10.1}<M/M_{\odot}<10^{11.2}$
(squares, Fig.~7{\it a}), B-band luminosities
$10^{10.5}<L_{\rm B}/L_{\rm B, \odot}<10^{11.3}$
(filled circles, Fig.~7{\it b}) and
$10^{10.0}<L_{\rm B}/L_{\rm B, \odot}<10^{10.5}$
(filled squares, Fig.~7{\it b}), and \citet[][Z07]{zheng07} with
$10^{10.0}<M/M_{\odot}<10^{10.5}$
(diamonds, Fig.~7{\it a}).
In these samples the stellar masses are roughly
comparable with our simulated ones. To convert the SSFR into $L_{\rm X}/M$, we
have made use of the local $L_{\rm X}-$SFR relation derived by
\citet{persic07}.

It seems that our simulated $\log(L_{\rm X}/M)$ vs. $z$
relations match the observations quite well.
Our calculations reveal that the modeled stellar masses
are similar to each other within 10\% when we use
different types of IMF, suggesting that the variation of the $L_{\rm X}/M$ ratio
is mainly caused by the differences in X-ray luminosity $L_{\rm X}$.
The values of $L_{\rm X}/M$ can vary by a factor of $\sim$7
between models (KTG93 vs. BG03), as seen in Fig.~7$a$.

The nearly constant values of $L_{\rm X}/L_{\rm B}$ in Fig.~7$b$
seem not properly  match the observed increase in $L_{\rm X}/L_{\rm
B}$ with $z$ \citep{hornschemeier02,lehmer08}. The discrepancy
originates from the fact that in our simulations we have roughly
$L_{\rm X}\propto L_{\rm B}$, giving a  flat
$L_{\rm X}/L_{\rm B}-z$ relation, while observationally it was found
that $L_{\rm X}\propto L_{\rm B}^{1.5}$ (Shapley et al. 2001; Fabbiano \&
Shapley 2002; Lehmer et al. 2008), leading to increasing $L_{\rm
X}/L_{\rm B}$ with $z$. Fabbiano \& Shapley (2002) have discussed
the X-ray-B-band luminosity correlation, and suggested that the nonlinear
power law dependency in disk galaxies is likely to be due to extinction in dusty
star-forming regions, which attenuates light from the B-band more
effectively than it does in the X-ray band. This means that the intrinsic
B-band luminosity calculated here is generally larger than the measured
one, which suffers local extinction in the galaxies. Supporting evidence for
this hypothesis also comes from the strong positive correlations
between $L_{\rm X}/L_{\rm B}$ and the ultraviolet dust-extinction
measure $(L_{\rm IR}+L_{\rm UV})/L_{\rm UV}$, and  the correlation
between $L_{\rm X}/L_{\rm B}$ and IR color \citep{lehmer08}. Thus,
if the increase in $L_{\rm X}/L_{\rm B}$ with $z$ in the late-type
galaxies is due to an increase in their star formation activity
\citep{fabbiano02,lehmer08}, our results are compatible with the
observational data at least qualitatively.


Figure~8 shows the corresponding cumulative X-ray
luminosity functions. In general they are similar to those in the case of
constant SF.

Our simulations are obviously subject to many uncertainties. For
current population synthesis investigations it is difficult to
tell confidently which parameter combinations are the best or most
realistic by comparison with observations, since there are many
uncertainties in the (both explicit and implicit) assumptions and
input parameters, and simplifications in the treatment of the
detail evolutionary processes. For example, the simulated X-ray
luminosity $L_{\rm X}$ depends on the adopted values of several
parameters, such as the bolometric correction factor $\eta_{\rm
bol}$, the common envelope efficiency parameter and so on, which
may alter its value severalfold. This further affects the values
of $L_{\rm X}/M$ and $L_{\rm X}/L_{\rm B}$ since the stellar mass
$M$ and the optical luminosity $L_{\rm B}$ is not sensitive to
these parameters. Despite these limitations it has become possible
to investigate the overall evolution of stars in galaxies with
population synthesis, and to draw useful information by comparing
theoretical predictions with observations.

\section{SUMMARY}

We have used an EPS code to calculate the X-ray evolution of
late-type galaxies, to investigate the relations between the X-ray
luminosity and other physical properties (i.e., optical luminosity,
stellar mass, and star formation history) of the galaxies, and how
these relations are influenced by the input parameters of star
formation and evolution (e.g., SFH, IMF, metallicity, and common
envelop efficiency, etc). The results are compared with multi-wavelength
analyses and observations of late-type galaxies
\citep{Shapley01,zheng07,lehmer08}.  In different cases of SF,
we find a common feature of decreasing X-ray
luminosity-to-mass ratio $L_{\rm X}/M$ with time, in agreement with
observations (Fig.~7$a$). We show that the decrease of $L_{\rm
X}/M$ results from slow evolution of $L_{\rm X}$ of XRBs and the stellar mass
accumulation with time in galaxies, without requiring that lower mass
galaxies have higher SSFR than more massive ones suggested before (Brinchmann et al.
2004; Bauer et al. 2005; Noeske et al. 2007a; Feulner et al. 2005;
Zheng et al. 2007). The $L_{\rm X}/L_{\rm B}$ ratios in all cases
rise rapidly in the first $\sim 10^8$ yr to $\sim 10^{30}$ erg
s$^{-1}$$L_{\rm B,\odot}^{-1}$, then stay nearly constant afterward
for a given model, and are not sensitive to the SFH details in
the galaxies (see Fig.~7$b$). This seems to be in conflict with the
observed increase in $L_{\rm X}/L_{\rm B}$ with $z$
\citep{hornschemeier02,fabbiano02,lehmer08}. The discrepancy may be
due to different obscured star formation activities in galaxies at
higher redshifts \citep{fabbiano02,lehmer08}. This will be investigated
by future high-resolution X-ray and optical observations of galaxies
at high redshifts.

\acknowledgments We thank Bret Lehmer and Xian-Zhong Zheng for
providing the relevant data plotted in Fig.~7 and helpful
suggestions, and an anonymous referee for detailed and constructive
comments. This work was supported by the Natural Science Foundation
of China (under grant number 10873008), the National Basic Research
Program of China (973 Program 2009CB824800), the National Natural
Science Foundation (under grant number 11003005), the Fundamental
Research Funds for the Central Universities, Jiangsu Project
Innovation for PhD candidates (0201001504) and National High
Performance Computing Center (at Xi'an).

\newpage

\begin{table}
\caption{Parameters adopted for each model. Here $\alpha_{\rm CE}$
is the CE efficiency parameter, $q$  the initial mass ratio, IMF the
initial mass function, $f$ binary fraction, $Z$ metallicity in
solar units, and SFD the duration of star formation in the
simulated galaxy.} \centering

\begin{tabular}{c|ccccccc}\hline\hline
& Model & $\alpha_{\rm CE}$ & P(q)    & IMF &  $f$  &  $Z$  &  SFD  \\
&   &   &    &     &     &  $Z_{\odot}$ & (Myr)\\   \hline
\multirow{9}{*}{Constant SF}
  &    M1 & 0.3 & $\propto q^{0}$ & KROUPA01     &   0.5 & 1.0  & 14000\\
  &    M2 & 1.0 & $\propto q^{0}$ & KROUPA01     &   0.5 & 1.0  & 14000\\
  &    M3 & 0.3 & $\propto q^{1}$ & KROUPA01     &   0.5 & 1.0  & 14000\\
  &    M4 & 0.3 & $\propto q^{0}$ & KROUPA01     &   0.8 & 1.0  & 14000\\
  &    M5 & 0.3 & $\propto q^{0}$ & KTG93        &   0.5 & 1.0  & 14000\\
\cline{2-8}
  &    M6 & 0.3 & $\propto q^{0}$ & KROUPA01     &   0.5 & 1.5  & 14000\\
  &    M7 & 0.3 & $\propto q^{0}$ & KROUPA01     &   0.5 & 0.5  & 14000\\
  &    M8 & 0.3 & $\propto q^{0}$ & KROUPA01     &   0.5 & 0.1  & 14000\\
  &    M9 & 0.3 & $\propto q^{0}$ & KROUPA01     &   0.5 & 0.02  & 14000\\ \hline
\multirow{3}{*}{Star-burst SF}
  &    M10& 0.3 & $\propto q^{0}$ & KROUPA01     &   0.5 & 1.0  & 100\\
  &    M11& 0.3 & $\propto q^{0}$ & KROUPA01     &   0.5 & 1.0  & 20\\
  &    M12& 0.3 & $\propto q^{0}$ & KROUPA01     &   0.5 & 0.02 & 100\\ \hline
 \hline
\end{tabular}
\end{table}

\begin{figure}
  \centering
  \includegraphics[width=12cm,height=15cm]{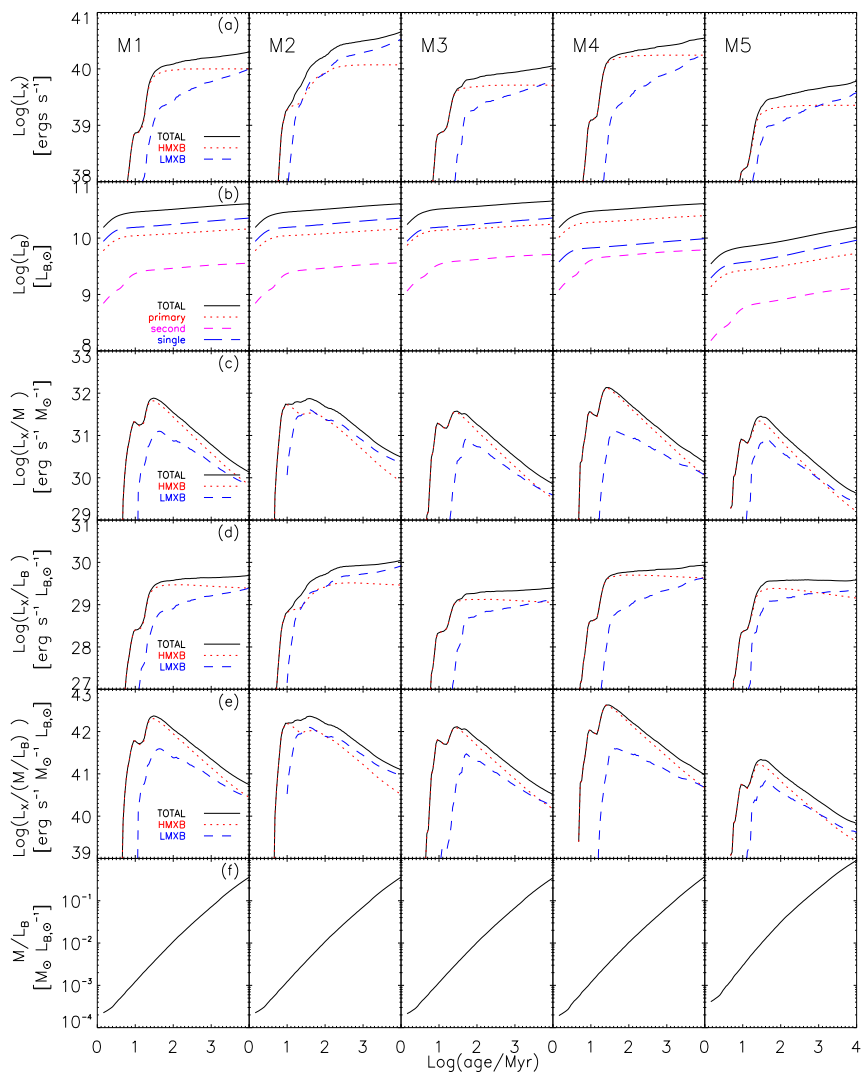}
\caption{The evolution of ($a$) the X-ray luminosity ($L_{\rm
X}$), ($b$) the optical luminosity ($L_{\rm B}$), ($c$) the
X-ray luminosity-to-stellar mass ratio ($L_{\rm X}/M$), ($d$) the X-ray-to-B
band luminosity ratio ($L_{\rm X}/L_{\rm B}$), ($e$) the $L_{\rm
X}/(M/L_{\rm B})$ ratio, and ($f$) the stellar-mass-to-B band
luminosity ratio ($M/L_{\rm B}$) with time in the constant star
formation case. Here the metallicity is $Z_{\odot}$. We assume
that the secondary mass distribution follows the power-law
$P(q)=q^{\alpha}$ and the binary fraction is $f$. The left panels
show the results in the basic model with $\alpha_{\rm CE}=0.3$,
$\alpha=0$, $f=0.5$ and KROUPA01 IMF. The other models from left
to right are with $\alpha_{\rm CE}=1.0$, $\alpha=1$, $f=0.8$ and
KTG93 IMF, respectively.}
  \label{Fig. 1}
\end{figure}

\begin{figure}
  \centering
  \includegraphics[width=15cm,height=9cm]{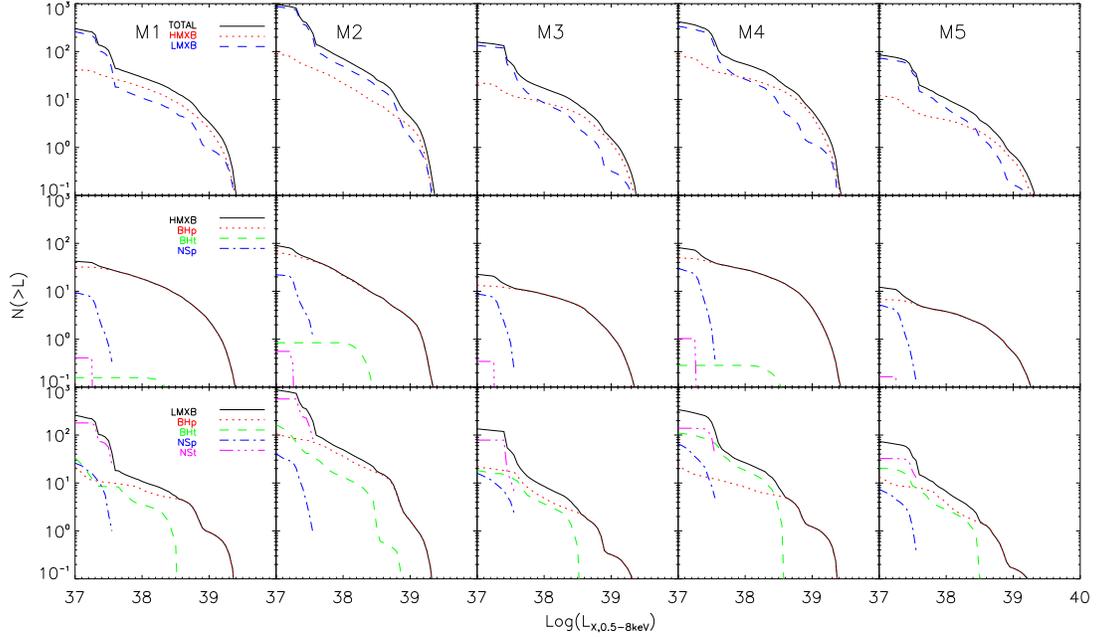}
\caption{The cumulative X-ray luminosity functions (Top: HMXBs +
LMXBs; Middle: HMXBs; Bottom: LMXBs). The model parameters are the
same as in Fig.~1. The dotted, dashed, dash-dotted and
dash-dot-dotted lines in middle and bottom panels represent black
hole persistent (BHp) and transient (BHt), neutron star persistent
(NSp) and transient (NSt) sources, respectively.}
  \label{Fig. 2}
\end{figure}

\newpage

\begin{figure}
  \centering
  \includegraphics[width=15cm,height=18cm]{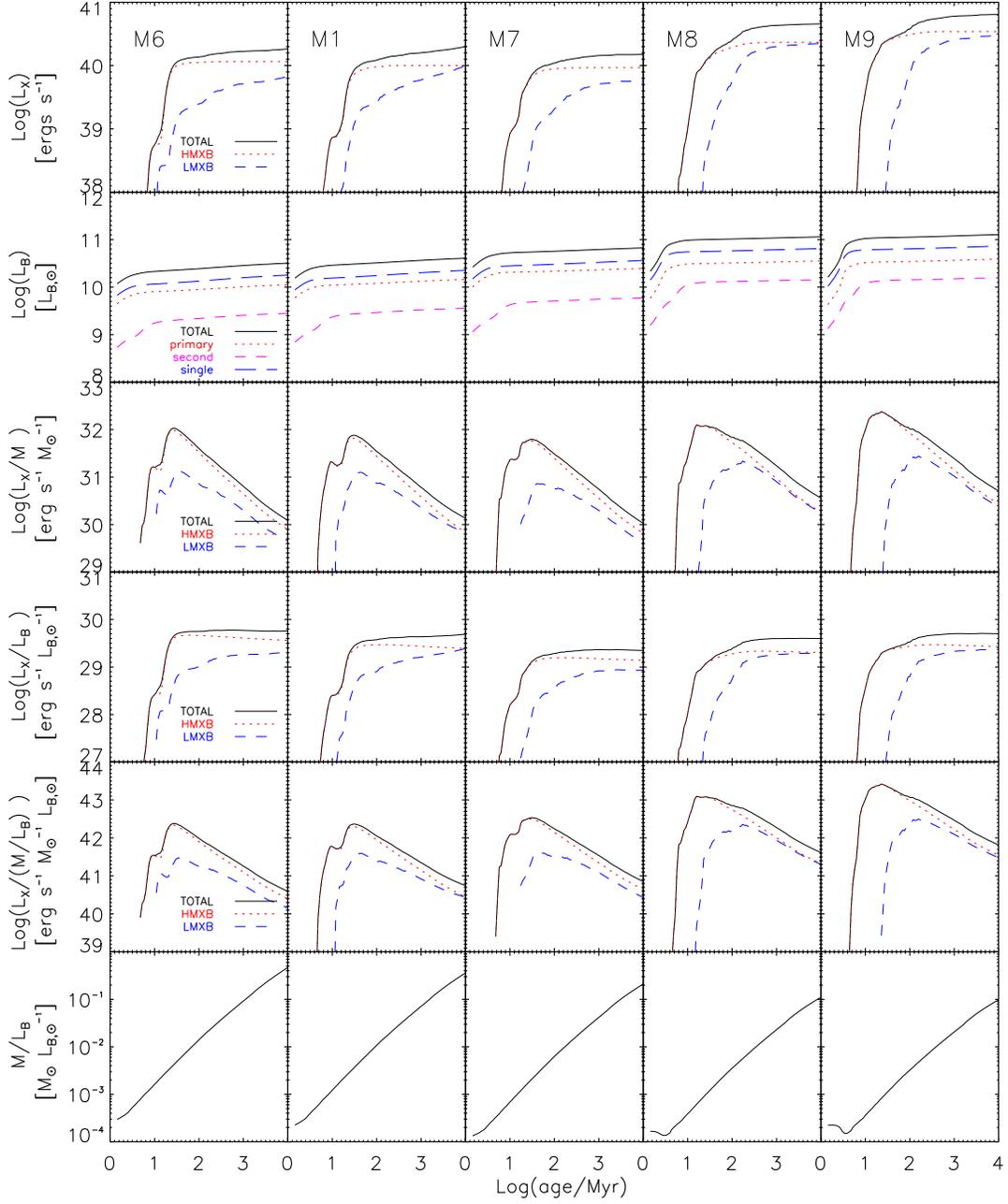}
\caption{Same as Fig.~1 but for different metallicities, which are taken
to be  $1.5 Z_{\odot}$, $Z_{\odot}$, $0.5 Z_{\odot}$, $0.1 Z_{\odot}$, and
$0.02 Z_{\odot}$ from left to right, corresponding to models M6, M1,
M7-M9,  respectively.}
  \label{Fig. 3}
\end{figure}


\begin{figure}
  \centering
  \includegraphics[width=15cm,height=9cm]{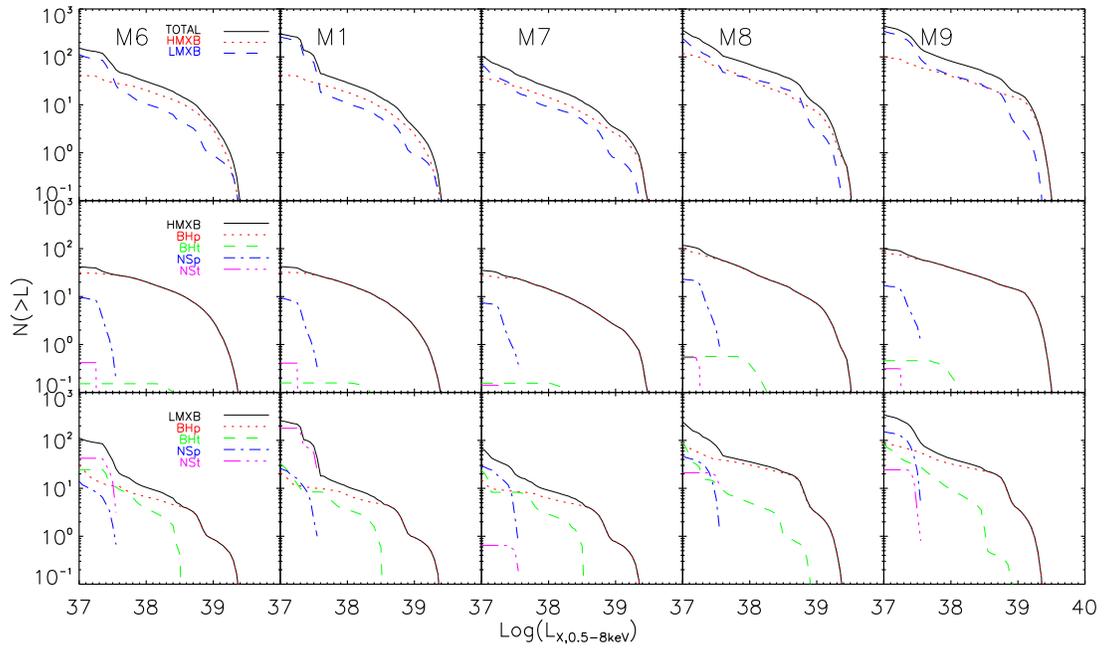}
 \caption{The cumulative X-ray luminosity functions (Top: HMXBs + LMXBs;
Middle: HMXBs; Bottom: LMXBs). The model parameters are the same as in
Fig.~3.}
  \label{Fig. 4}
\end{figure}

\begin{figure}
  \centering
  \includegraphics[width=9cm,height=18cm]{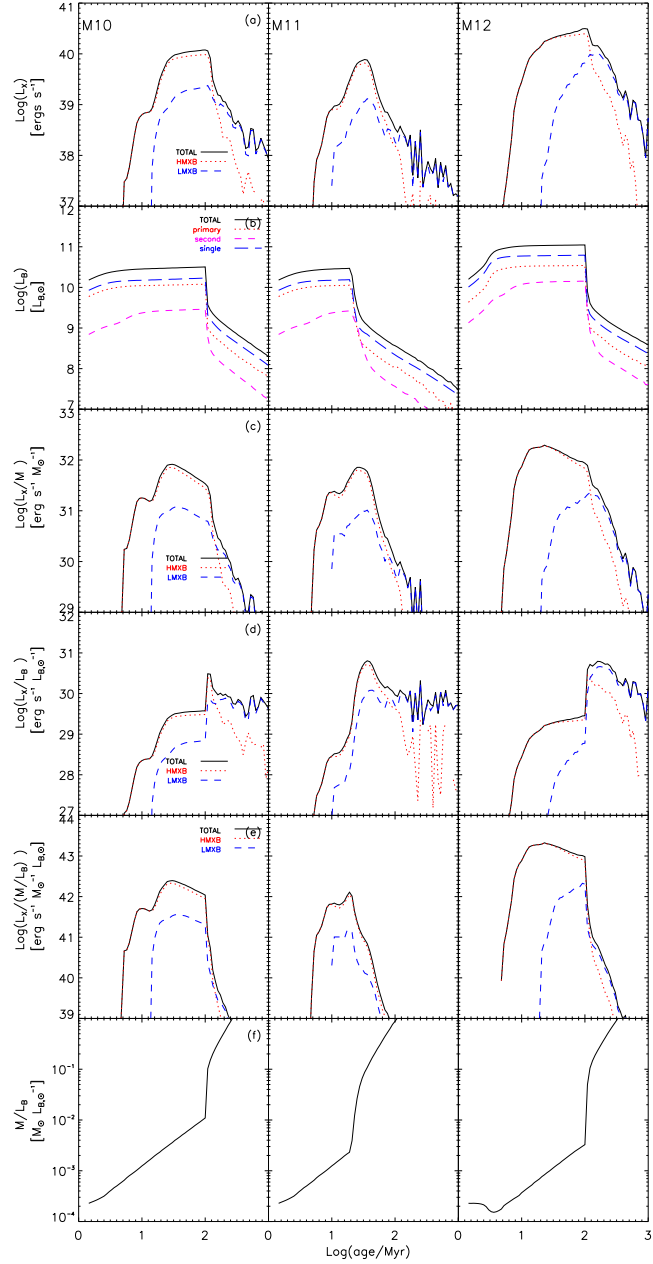}
\caption{The $L_{\rm X}$, $L_{\rm B}$, $L_{\rm X}/M$, $L_{\rm
X}/L_{\rm B}$, $L_{\rm X}/(M/L_{\rm B})$, $M/L_{\rm B}$  evolution
with time in the star-burst case.  Here
the metallicities and SFH are $Z_{\odot}$/100 Myr (left, M10),
$Z_{\odot}$/20 Myr (middle, M11), and $0.02 Z_{\odot}$/100 Myr (right, M12),
respectively.}
  \label{Fig. 5}
\end{figure}

\begin{figure}
  \centering
\includegraphics[width=10cm,height=15cm]{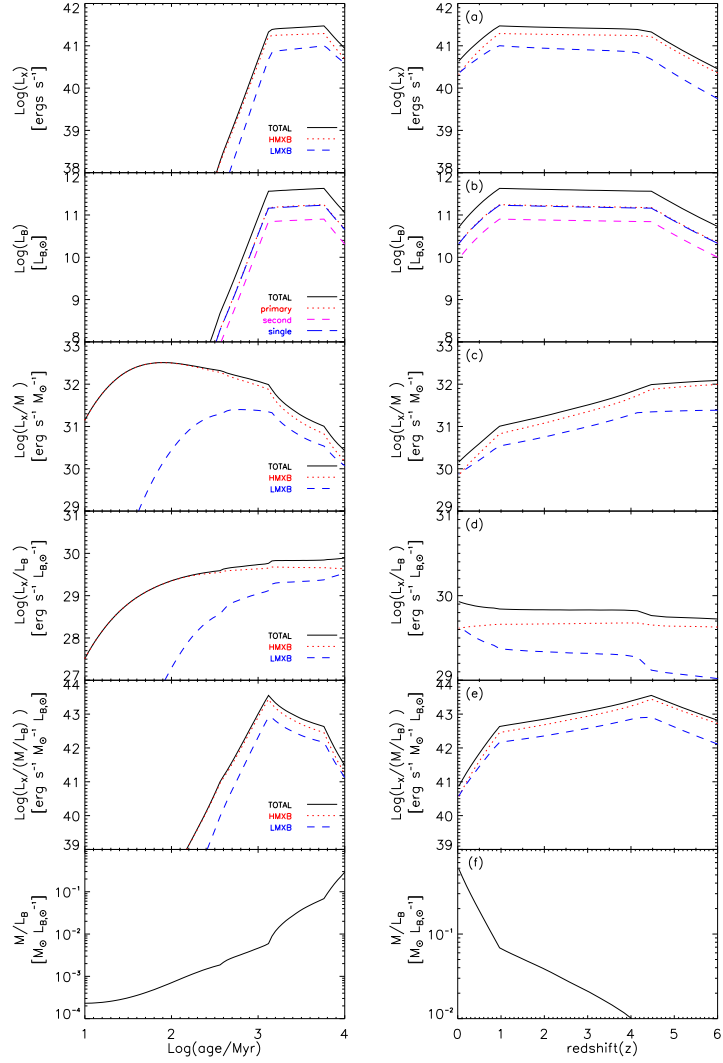}
\caption{The $L_{\rm X}$, $L_{\rm B}$, $L_{\rm X}/M$, $L_{\rm
X}/L_{\rm B}$, $L_{\rm X}/(M/L_{\rm B})$, $M/L_{\rm B}$
evolution with time (left) and
redshift $z$ (right), respectively. Here we have assumed a cosmic
star formation history \citep[from][]{hopkins06} and a cosmic
metallicity evolution history \citep[from][]{langer06}.}
  \label{Fig. 6}
\end{figure}

\begin{figure}
  \centering
 \includegraphics[width=15cm,height=7.5cm]{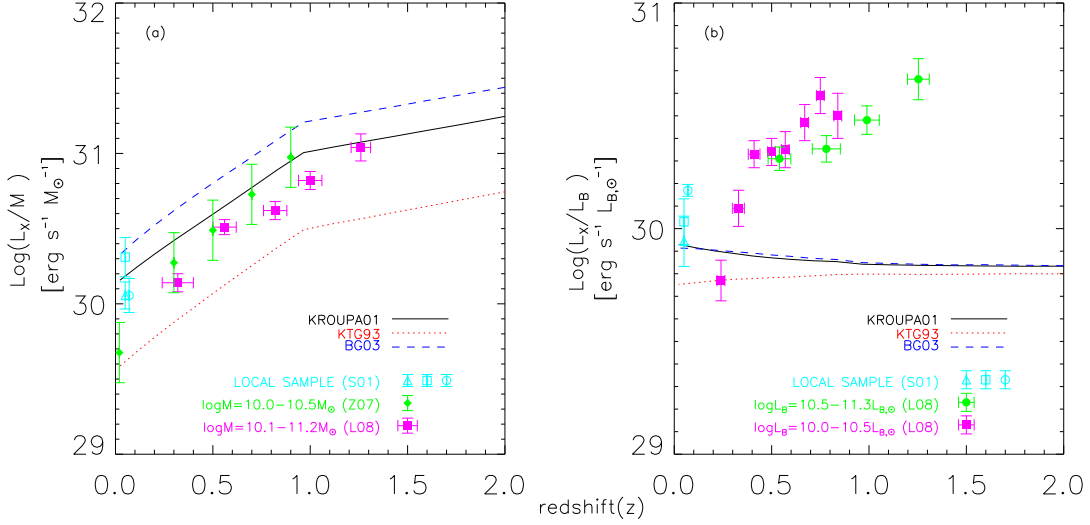}
 \caption{Same as panels $c$ and $d$ in Fig.~6 but enlarged
with $z=0-2.0$ for comparison
 with observations. The solid, dotted and dashed line represent
 modeled results with IMFs of Kroupa (2001, KROUPA01), Kroupa, Tout \&
 Gilmore (1993, KTG93), and Baldry \& Glazebrook (2003, BG03),
 respectively. Also shown are the measured values of $\log(L_{\rm X}/M)$
 (left panel) selected by $M$ ($10^{10.1}<M/M_{\odot}<10^{11.2}$,
squares), and $\log(L_{\rm X}/L_{\rm B})$ (right panel) selected by
$L_{\rm B}$ ($10^{10.5}<L_{\rm B}/L_{\rm B, \odot}<10^{11.3}$,
filled circles; $10^{10.0}<L_{\rm B}/L_{\rm B, \odot}<10^{10.5}$,
filled squares), respectively, for stacked normal late-type galaxy
samples derived by \citet[][L08]{lehmer08}. The
symbols are the same as  in
Fig.~10 in L08. We converted the SSFR to $L_{\rm X}/M$ in
\citet[][Z07]{zheng07} samples of the
corresponding stellar mass bin ($10^{10.0}<M/M_{\odot}<10^{10.5}$,
diamonds), using the local $L_{\rm X}-$SFR relation derived by
\citet{persic07}. The data for normal late-type galaxies in the local universe
(open symbols) are from the Shapley et al.
(2001, S01) samples. 
}
  \label{Fig. 8}
\end{figure}

\begin{figure}
  \centering
 \includegraphics[width=15cm,height=5cm]{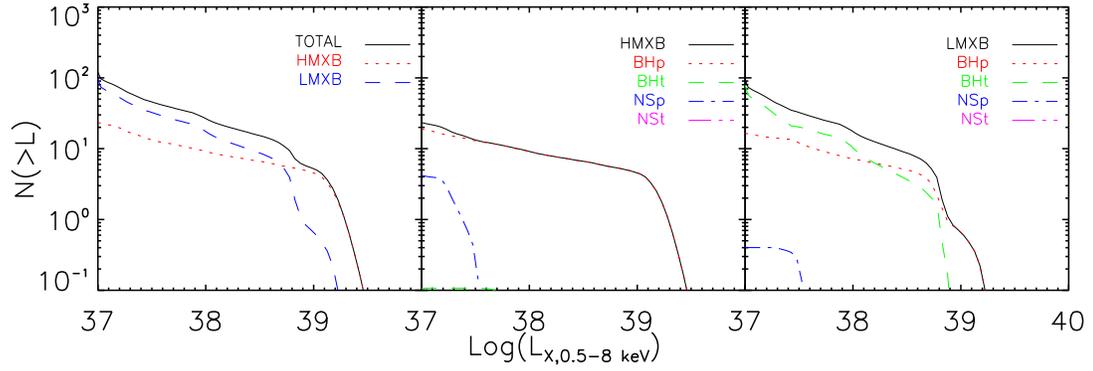}
 \caption{The cumulative X-ray luminosity function (Left:H+L;
Middle:H; Right: L). Here we have assumed a cosmic star formation
history \citep[from][]{hopkins06} and a cosmic metallicity
evolution history \citep[from][]{langer06}.}
  \label{Fig. 7}
\end{figure}

\end{document}